\documentclass{aastex631}

\usepackage{amsmath}
\newcommand{\E}{\mathbb{E}}

\shorttitle{Physical quantities to generate and retrieve solar magnetic active regions}
\shortauthors{Chatterjee et al.}

\graphicspath{{./}{figures/}}

\begin{document}

\title{Deep Generative model that uses physical quantities to generate and retrieve solar magnetic active regions}

\author[0000-0002-5014-7022]{Subhamoy Chatterjee}
\affiliation{Southwest Research Institute, Boulder, CO 80302, USA}

\author[0000-0002-4716-0840]{Andr\'es Mu\~noz-Jaramillo}
\affiliation{Southwest Research Institute, Boulder, CO 80302, USA}

\author[0000-0001-5066-8509]{Anna Malanushenko}
\affiliation{High Altitude Observatory, NCAR, Boulder, CO, USA.}

\begin{abstract}

Deep generative models have shown immense potential in generating unseen data that has properties of real data. These models learn complex data-generating distributions starting from a smaller set of latent dimensions.  However, generative models have encountered great skepticism in scientific domains due to the disconnection between generative latent vectors and scientifically relevant quantities. In this study, we integrate three types of machine learning models to generate solar magnetic patches in a physically interpretable manner and use those as a query to find matching patches in real observations. We use the magnetic field measurements from Space-weather HMI Active Region Patches (SHARPs) to train a Generative Adversarial Network (GAN). We connect the physical properties of GAN-generated images with their latent vectors to train Support Vector Machines (SVMs) that do mapping between physical and latent spaces. These produce directions in the GAN latent space along which known physical parameters of the SHARPs change. We train a self-supervised learner (SSL) to make queries with generated images and find matches from real data. We find that the GAN-SVM combination enables users to produce high-quality patches that change smoothly only with a prescribed physical quantity, making generative models physically interpretable.  We also show that GAN outputs can be used to retrieve real data that shares the same physical properties as the generated query.  This elevates Generative Artificial Intelligence (AI) from a means-to-produce artificial data to a novel tool for scientific data interrogation, supporting its applicability beyond the domain of heliophysics.

\end{abstract}


\section{Introduction} \label{sec:intro}
Modern astronomical observatories produce hundreds of petabytes of data during their lifetime. Manually labeling and sifting through such a scale of data is becoming impossible in a human lifetime. A bigger problem is to process these data and retrieve information hidden in those large datasets. Self-Supervised-Learning (SSL)\citep{chen2020} has emerged as an effective machine learning approach to retrieve information by querying large unlabeled datasets. While it is easy to use an existing image as a query, sometimes the researcher may wish to find data for which there is no readily available physical reference. For example, Rogue active regions with unusual size, tilt, and location have been found to make substantial impacts on solar cycles \citep{Petrovay_Nagy_2018, Nagy_2017}. Such regions are rare in occurrence, and a scientist might want to create more examples of such regions with arbitrary configurations and incorporate them in a dynamo simulation to study varied impacts on solar cycles. This is further supported by a recent study \citep{pal2023} investigating the effect of anomalous active region configurations on the build-up of magnetic dipole moments with a surface flux transport model. Note that the definition in \citet{pal2023} only considers a subset of the possible anomalous active regions, since they model anomalous regions as simple bipolar magnetic regions with anti‑Joy tilts and/or anti‑Hale polarity. In contrast, one may also be interested in anomalous active regions that are unusually large in size and unsigned flux, with greater magnetic complexity (e.g. $\beta\gamma\delta$ as defined in \citet{Kondrashova2023}), while still exhibiting anti‑Joy tilts and anti‑Hale polarity. Also, another study by \citet{jha2025} uses sunspot number cycle, along with the statistical distribution of physical properties, to create a synthetic bipolar active region catalog over solar cycle 1-24 and feed the same into the Advective Flux Transport simulation for reconstructing the solar surface magnetic field. The ability to generate synthetic data with desired physical characteristics and additionally use that as search query may greatly enhance scientific return from any dataset in question.

Generative Artificial Intelligence (AI) models use complex architectures to learn data-generating distributions and have been highly successful in producing synthetic data with relevant characteristics similar to those in real data.  This makes them ideal for empowering users to find structures efficiently without sifting through all data. In this work we show how Generative AI created data can be used as a query for SSLs to retrieve matches from real data. As part of this process, we show how Generative AI has to be modified so that it can create data based on physical quantities.

One of the main challenges of using Generative AI in scientific applications is the obscure nature of hidden data-generating vectors within the model, also known as latent vectors. In Fourier space, for example, an image is uniquely represented by a set of coefficients in the space of the basis functions, which, in turn, correspond to different spatial frequencies, therefore the Fourier dimension space makes physical sense. But in the space that Generative AI operates in, the components of a latent vector that uniquely determines an image are non-interpretable, and also they interact with each other in a non-linear manner during the transformation that creates the image. 

Our contribution focuses on addressing the lack of physicality in Generative AI that acts as a barrier to its adoption as a scientific tool. We use supervised learning to map desired physical characteristics related to the obscure latent spaces of generative models, enabling the use of physical quantities to generate or modify physical queries and find real matching observations. This addresses a need that will only become worse as more astronomical data are produced and elevates Generative AI from a means-to-produce artificial data to a novel tool for scientific data interrogation.

Our approach can be easily applied to many astronomical datasets, such as emulating exoplanet transients and sifting through a large set of light curves for similar characteristics or generating astronomical objects (such as galaxies and nebulae) and finding similar examples in wide-field surveys. 

The focus of our study is solar magnetic active regions (ARs).  ARs are the major source of space weather events that impact Earth and drive variability in the solar system. As such, they are of major interest in solar physics.  We use a set of physical parameters, used in numerous studies in Heliophysics to describe and study the properties of ARs.

Approaches exist that convert AR parameters to AR images and use those for data-driven simulations. For example, simulations of the solar dynamo use ARs that are modeled as idealistic, simple bipolar structures (e.g. two Gaussian spots). However, such simple bipolar structures lack the properties of real ARs such as shape, texture, etc. In this study, we make use of a state-of-the-art approach \citep{shen2019} utilizing Generative Adversarial Networks (GANs) \citep{goofellow2014} in a supervised manner to modify generated images along directions that correspond to known AR parameters.

\begin{figure}[t!]
\hspace{-0.0\textwidth}
\includegraphics[width=1.0\linewidth]{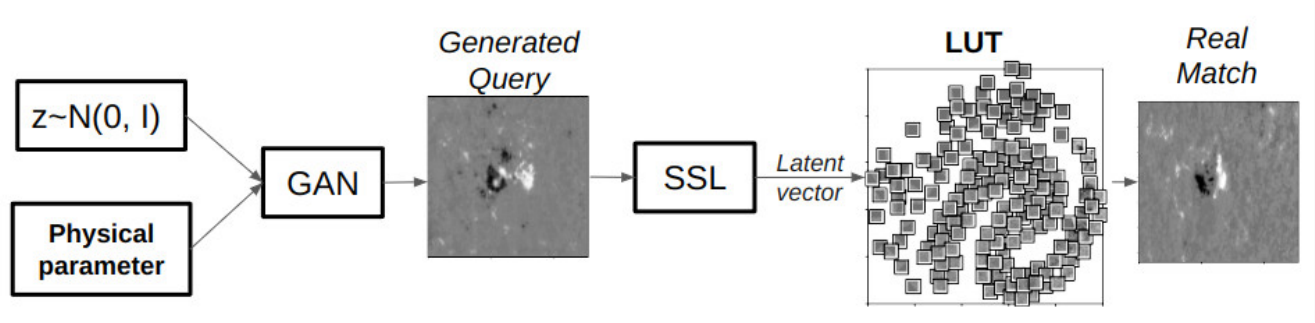}
\caption{\textbf{Graphical illustration of the designed and implemented solar image generation and retrieval pipeline.} The process involves a generative model (GAN), supervised learning to utilize physical parameter space, and self-supervised learning (SSL) for reverse real image-search. The Look-Up-Table (LUT) describes the mapping between an image and SSL-derived latent representation. }
\label{fig1}
\end{figure}

We design a pipeline (see Figure~\ref{fig1}) that uses GAN-generated data as a query and utilizes a SimSiam \citep{chen2020} SSL model to retrieve matches from observed data (which in our case are solar magnetic active regions). Development of this pipeline and results are presented in the remainder of the paper. We start by describing the observational data in \S\ref{sec:data}, and the methodology in \S\ref{sec:methods}, illustrating different components of the pipeline, namely the generative, supervised, and self-supervised models. We elaborate on the findings containing training results, physical manipulation of the generated images, and nearest neighbor search performance in \S\ref{sec:results}. We finally discuss the results and conclude in \S\ref{sec:discon}.

\newpage
\section{Data}\label{sec:data}
We use Space-weather HMI Active Region Patches (SHARPs)\citep{bobra2014} for this work. We acquire 49,552 patches of radial magnetic field on the solar surface, which corresponds to time sequences of \textbf{1839} unique patches from May, 2010 to July, 2023 in 12 minute cadence. We use the \texttt{Quality} flag to get rid of bad samples. The patches are in cylindrical equal area projection. It must be noted that the SHARPs come in various sizes, and to make them ready as input to deep learning models we interpolate the patches to the same image size of $128 \text{~pix} \times 128 \text{~pix}$. As more than 97\% of the images were bigger than the size $128\times128$ the resizing was done by area filtering getting rid of the aliasing effect, and the rest ($<3\%$) of images (smaller than $128\times128$) were resized through bilinear interpolation. We note that this upsampling operation with bilinear interpolation potentially causes smoother gradients and weaker fields than they would be on an image of original size 128$\times$128. 

We also scale the value of magnetic field, dividing by 1000 G, and clipping values to the range $[-1, 1]\times 1000\mbox{ G}$. It should be noted that although absolute field strengths greater than 1000G occurs in only 5\% of pixels in the majority of the images, the clipping results in saturation of strong AR regions with sunspots. This effect is partially mitigated by the resizing operation and the fact that we primarily used integral image properties to build the physical space. However, it must be noted that our end goal is not the produce the most realistic synthetic AR images. Instead we aimed to build a framework that connects GAN latent space to physical space and that can be used across GAN architectures or other generative models. To get more realistic synthetic ARs from GAN the clipping range can be increased and a nonlinear transformation (e.g. \texttt{Tanh}) can be applied to mitigate the effect of outliers.

\section{Methods}\label{sec:methods}

\textbf{GAN.} GAN\citep{goofellow2014} works as an adversarial two-player game between two modules, namely generator and discriminator, to learn the data-generating distribution ($p_x$).   With training, the generator ($G$) becomes superior in generating unseen data ($G(z)$) from just a noise vector ($z$, sampled from a standard normal distribution $p_{z}$) that has properties of real data ($x$) and the discriminator ($D$) simultaneously improves in distinguishing generated data from real data. Training stops when the Nash equilibrium is reached, making the data-generating distribution indistinguishable from the generative distribution. The shape of $G(z)$ is the same as that of the real data $x$. $D$ outputs a probability with $D(x)$ signifying the probability that $x$ is a real data point. The optimization of $G$ and $D$ are performed as a two-step process per training iteration by minimizing binary cross-entropy losses $D_{loss}$ and $G_{loss}$ respectively as described by the following equations-
\begin{equation}
    G_{loss} = -\E_{z\sim p_{z}}\log D(G(z))
\end{equation}

\begin{equation}
    D_{loss} = -\frac{1}{2}[\E_{z\sim p_{z}}\log (1 - D(G(z))) + \E_{x\sim p_{x}}\log D(x)]
\end{equation}

$G_{loss}$ is aimed at strengthening the generator to fool the discriminator into thinking a generated image as a real image. Thus a target of 1 is chosen for the generated image in the binary cross-entropy loss that constitutes the $G_{loss}$. If the discriminator can identify that the generated image as a fake, the generator is trained to produce more realistic samples. $D_{loss}$ is aimed at strengthening the discriminator in being able to distinguish between real data and the synthetic one. It is defined as a binary cross-entropy loss with 1 being the target for real data and 0 being the target for the generated data.

 The architecture of $G$ is constructed by sequential usage of a \texttt{linear} and three \texttt{deconvolution} blocks. The \texttt{linear} block and the first two \texttt{deconvolution} blocks use \texttt{LeakyReLU} nonlinearity and the last one uses \texttt{Tanh} nonlinearity. All the \texttt{deconvolution} blocks use \texttt{ConvTranspose2d} operation with \texttt{stride} 2 and kernel size $3\times3$.  The architecture of $D$ is constructed by sequential usage of three \texttt{convolution} blocks and a \texttt{linear} layer. All the \texttt{convolution} blocks make use of \texttt{LeakyReLU} nonlinearity and a \texttt{Sigmoid} nonlinearity is used after the \texttt{linear} layer. All the \texttt{convolution} blocks use \texttt{Conv2d} operation with \texttt{stride} 2 and kernel size $3\times3$. We make the architecture choice of $G$ and $D$ to ensure training stability. 

 We train GAN on Space-weather HMI Active Region Patches (SHARPs)\citep{bobra2014}. The parameters of $D$ are not updated when $G$ is trained. When $D$ is trained the output of $G$ is used and its optimization is stopped by disconnecting it from the model graph. We use Adaptive Moment Estimation (ADAM) as the optimizer during training.
\\
\\
\textbf{SVM.} Support Vector Machines (SVMs)\citep{svm1995} fall under a class of supervised Machine Learning models that offer flexibility in dealing with high-dimensional datasets. They are used in both classification and regression. SVMs offer different kernel functions to draw decision boundaries for classification and use a subset of training data (also known as `support vectors') to derive the decision boundary. We train a set of linear Support Vector Machines (SVMs) to learn the decision boundaries in the latent space that separate regimes of higher and lower values of physical parameters related to magnetic patches.
\\
\\
\textbf{SimSiam.} Simsiam\citep{chen2020} is a state-of-the-art self-supervised learning model that learns lower-dimensional representations (known as `latent vectors') of images such that the representations are invariant across a specified set of augmentations. It works by creating two augmented versions of an image and passing each through a backbone architecture shared by two branches. The backbone architecture is implemented by a Convolutional Neural Network -- ResNet. The outcome of the backbone is then passed through a projection head in both branches and a prediction head in one of the branches. Multilayer perceptrons implement the projection and prediction heads. The outcome of projection ($\vec{p}$) and prediction heads ($\vec{z}$) from two branches are then compared by a negative cosine similarity loss $L = -\frac{\vec{z}.\vec{p}}{||\vec{z}||||\vec{p}||}$. The model is trained by minimizing $L$. We train SimSiam on the SHARP data to learn 100-dimensional latent representations that are invariant on translation, zoom, flip, and a restricted range of rotation.

\section{Results}\label{sec:results}
The results from our designed and implemented pipeline (Figure~\ref{fig1}) are described sequentially in the following subsections.

\textbf{Training Generative Model.} We train a GAN to generate $128 \text{~pix} \times 128 \text{~pix}$ magnetic patches from a 100-dimensional noise vector ($z$; also referred to as `latent vector' throughout the paper) sampled from a standard normal distribution. As described before, (1) We implement the generator as a simple Convolutional Neural Network (CNN)-based decoder architecture and use $\tanh$ transformation as the last step of the generator architecture, thus making the generated images bounded within [-1, 1]. 
(2) We construct a discriminator, based on a CNN-based encoder architecture that we built from scratch, to classify the generated magnetic patches as `fake' and SHARP patches as `real'. The last layer in the discriminator is a sigmoid-function transformation.

\begin{figure}[t!]
\includegraphics[width=1.0\linewidth]{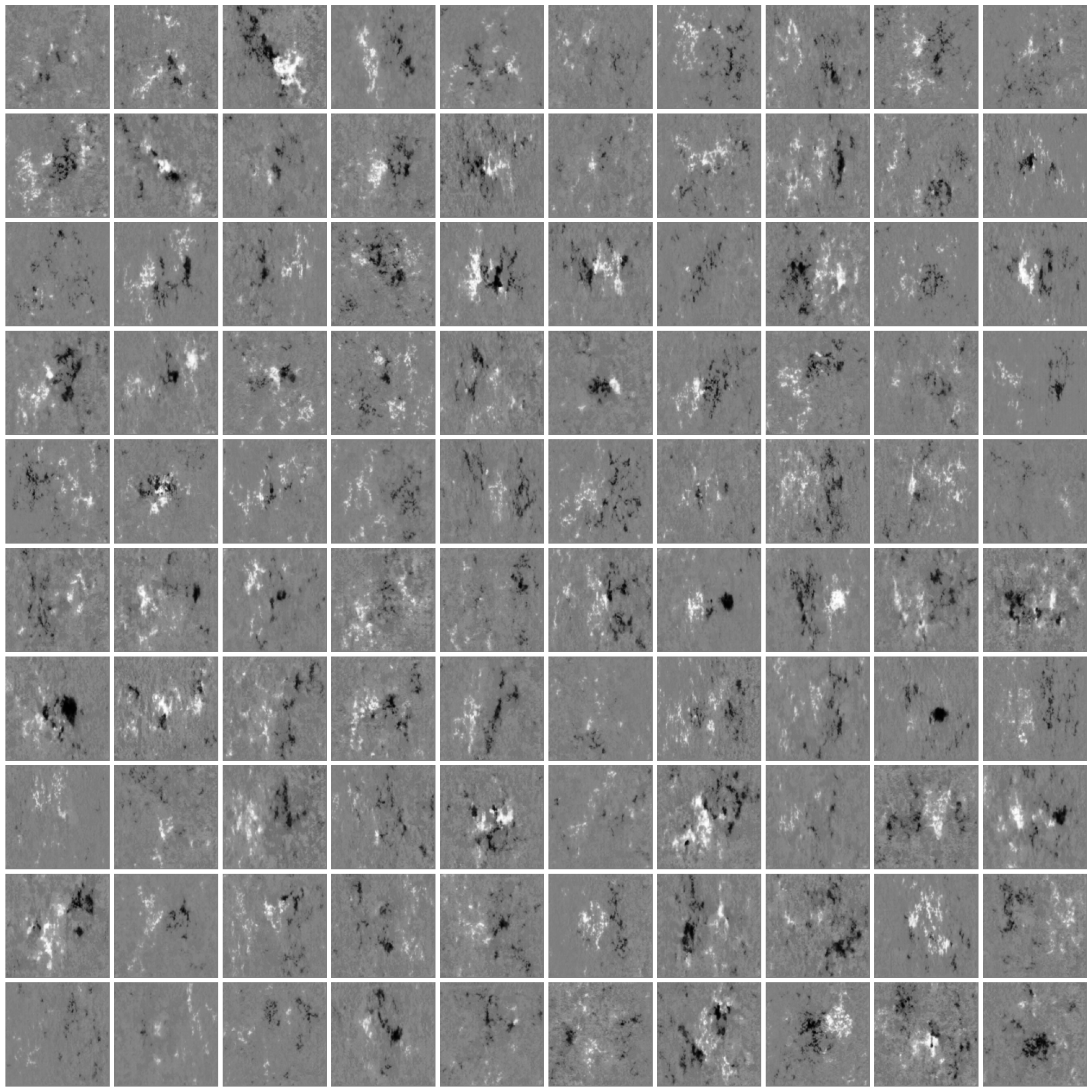}
\caption{\textbf{A zoo of GAN-generated magnetic Active Regions that depending on complexity play an important role in driving space weather events.} The shown AR images are generated using 100 latent vectors randomly sampled from a standard normal distribution. The images with diffused flux are seen as the temporal evolution of SHARPs is used during the training of the GAN.}
\label{fig2}
\end{figure}

We train the GAN for 200 epochs with a batch size of 32 and a learning rate of 0.0005. In the first 100 epochs of GAN training we don't observe any increasing  or decreasing trend in the $G_{loss}$ and $D_{loss}$. However, over epochs 100-200 we find a clear increasing trend in $G_{loss}$ and decreasing trend in $D_{loss}$ producing inferior quality of generated images. For further inference, we pick the epoch, from first 100 epochs, for which the discriminator shows the least accuracy in distinguishing generated data from real. Figure~\ref{fig2} illustrates the results from our GAN as a 10$\times$10 matrix of generated magnetic patches. We note that the flux in generated images ranges from low to high. This is because GAN is trained on time sequences of magnetic patches during their emergence and evolution.

\textbf{Interpretable Generation of Magnetic Patches.} Although the GAN can generate realistic solar magnetic patches, interpretable manipulation of the images using the embedding space is non-trivial. Gradually changing the value of each dimension in noise vector $z$ gives rise to smooth changes in the generated images, but those changes are often not physically interpretable, as multiple features tend to evolve simultaneously. To control the generation in a physically meaningful space we found it is most useful to use a supervised approach to learn decision boundaries in the latent space of GAN. The decision boundary, for a given physical parameter, separates low and high values of the same parameter calculated from GAN-generated images. This approach is illustrated in the following paragraphs.  

\begin{figure}[t!]
\centering
\hspace{-0.0\textwidth}
\includegraphics[width=0.9\linewidth]{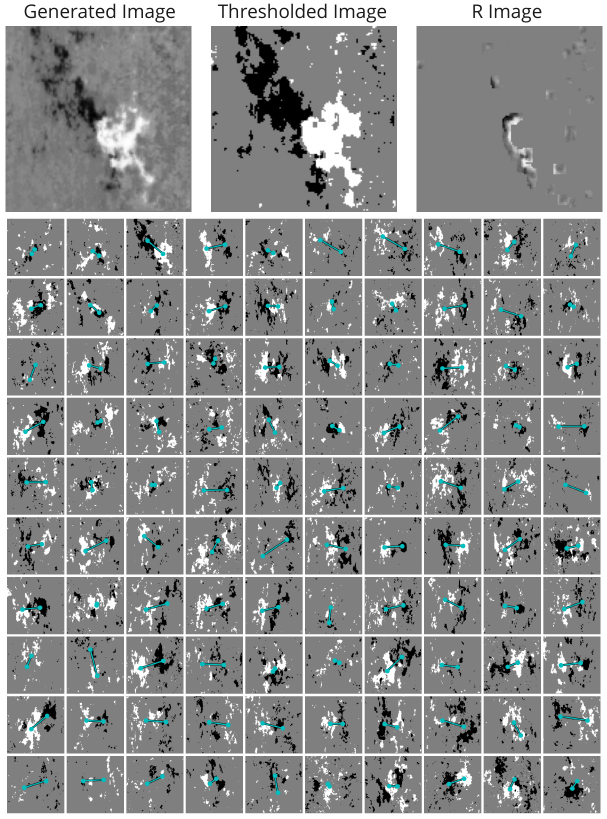}
\caption{\textbf{Calculation of physical parameters from thresholded GAN generated magnetic patches.} The top row shows a GAN-generated active region, its thresholded version, and the derived regions of polarity inversion with a high field gradient from left to right. The collage starting from the second row shows the thresholded version of all GAN-generated images in Figure~\ref{fig2} with field-strength-weighted centroids of positive and negative polarity connected by a cyan line.}
\label{fig3}
\end{figure}

We first generate 10,000 magnetic patches using the GAN from randomly sampled noise vector $z$. We then calculate physical parameters such as the total unsigned field strength($TUF$), pole-to-pole separation ($PSEP$), and the parameter $R$. This selection of parameters was primarily driven by their connection to space weather events such as flares \citep{Schrijver_2007, kvSande_2023}. $R$ represents a total unsigned field in subregions with a high gradient of the field with polarity transition (Figure~\ref{fig3}). Following \citet{Schrijver_2007}, we calculate $R$ by dilating the pixels with field strength above 150 G or below -150 G, and use the intersection of those regions as a mask for high gradient regions in the patches. The unsigned field is then integrated within these regions to get the value of $R$. For the images that do not have the occurrence of such a region, the R-value is estimated to be 0. Using the same threshold of $\pm150\mbox{ G}$ we also calculate $PSEP$ as the distance between field strength-weighted centroids of pixels within positive and negative polarity regions (shown as cyan lines in Figure~\ref{fig3}).  To calculate $TUF$ we sum unsigned field strength over all the image pixels.   

We make a lookup table of the 100-dimensional noise vector $z$ vs. the calculated parameters for all 10,000 generated images and use the median of each parameter as a threshold to assign a binary label. Using these binary labels we train a linear Support Vector Machine (SVM) to learn the decision boundary on the 100-dimensional latent space that separates regimes of higher and lower values of the selected parameters. We use 7,000 data points for training and 3,000 for validation. In the linear SVM classifier, we got the best results with a regularization strength of 1. We did not require any class reweighting as we used balanced ($50\%$ positive, $50\%$ negative) classes. We acquire the normal vectors ($\hat{\textbf{n}}^{p}$) for the decision boundaries of three trained SVMs ($p \in \{\text{TUF}, \text{PSEP}, \text{R}\} $).

\begin{figure}[t!]
\hspace{-0.0\textwidth}
\includegraphics[width=1.0\linewidth]{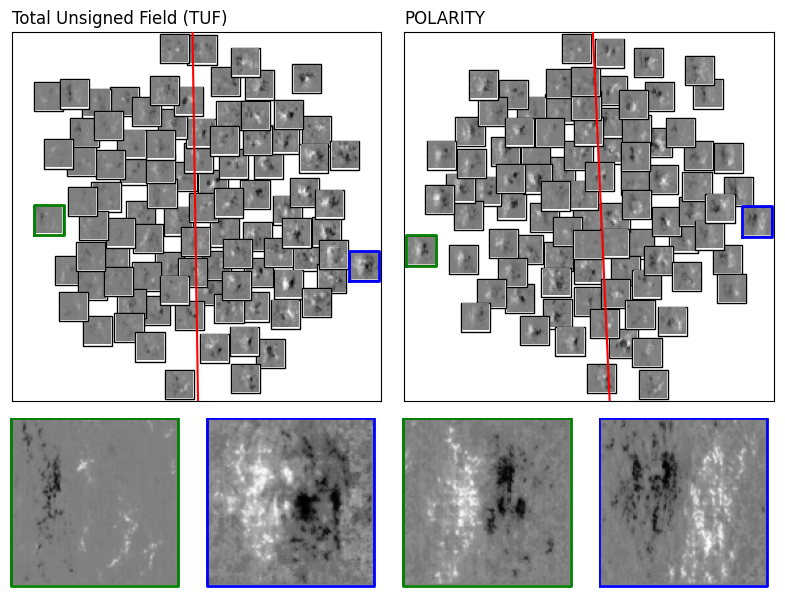}
\caption{\textbf{Generated solar images overlaid on the 2-dimensional projection of latent space.} The projection was performed such that the decision boundary hyperplane becomes a line (marked in red) in the depicted 2d latent space. The decision boundary marks a clear separation between the physical properties (total unsigned field (TUF), and polarity) represented by the generated images. The bottom rows show zoomed-in views (outlined with green and blue rectangles) of the generated images from both extremes across the decision boundaries.}
\label{fig4}
\end{figure}

To visualize the latent space and the SVM-derived decision boundary in 2-dimensions described by ($z_1^{\prime}, z_2^{\prime}$) we project the 100-dimensional space ($z_1, z_2, .., z_{100}$) using a simple transformation: $z_1^{\prime} = z_1$ and $z_2^{\prime}=\frac{1}{n_2}\sum_{i=2}^{100}n_iz_i$, where ($n_1, n_2, ..,n_{100}$) defines the normal ($\hat{\textbf{n}}^{p}$) to the hyperplane defining the decision boundary and $n_2\ne0$. This transformed the hyperplane to a line in the visualized 2-dimensions. In Figure~\ref{fig4}, one can observe a clear demarcation between physical parameters represented by images above and below the decision line.

\begin{figure}[t!]
\hspace{-0.0\textwidth}
\includegraphics[width=1.0\linewidth]{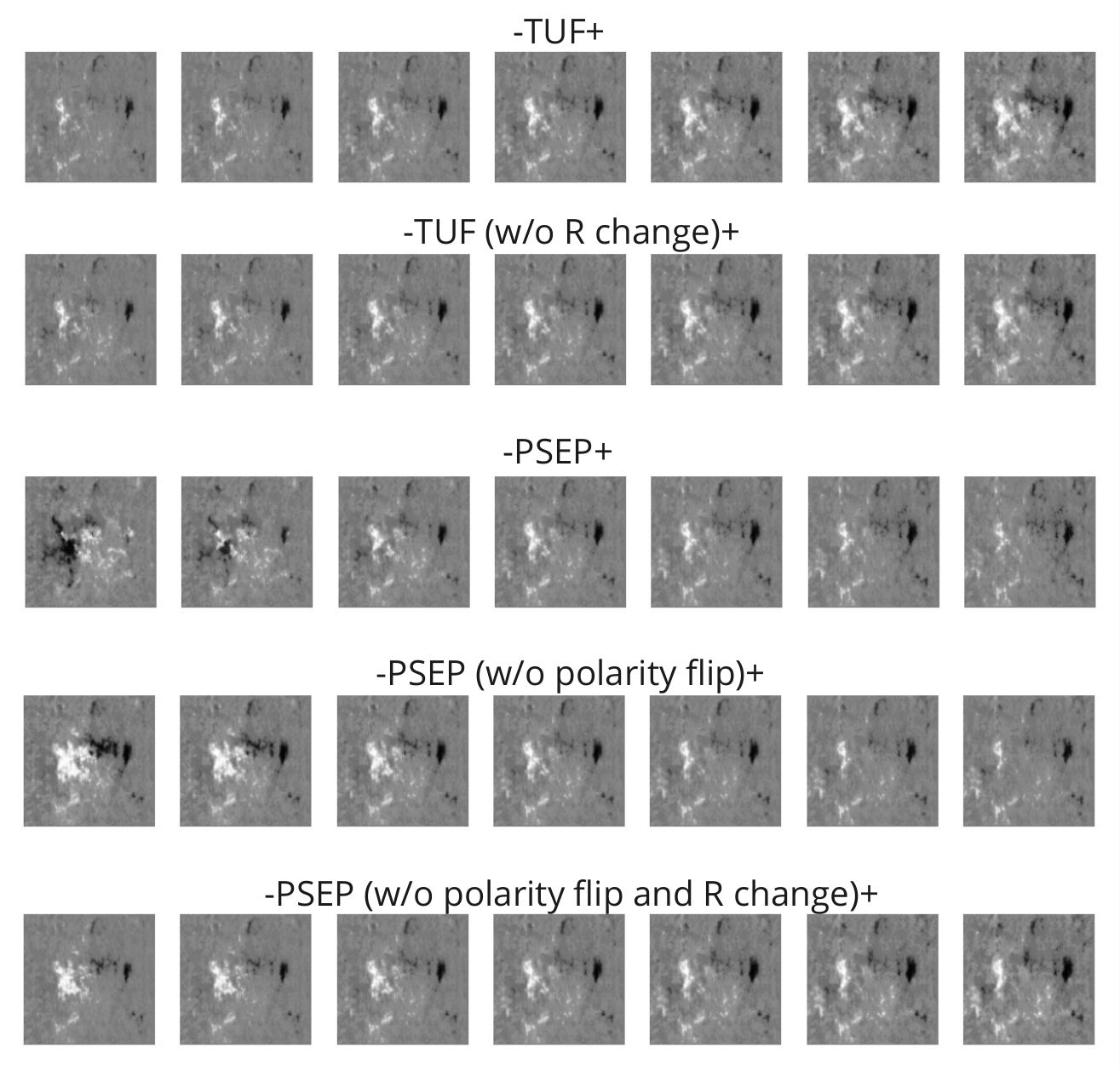}
\caption{\textbf{Manipulation of GAN generated image along two different physically interpretable directions}. Two directions are shown -- total unsigned field (TUF; top two rows) and polarity separation (PSEP; bottom three rows). The 2nd row from the top shows the result of decoupling TUF from the direction along which the R changes. The 4th row shows the result of decoupling PSEP from the direction along which polarity is flipped and the bottom-most row shows the result of decoupling PSEP from both polarity flip and change in R.}
\label{fig5}
\end{figure}

Given a latent vector ($\textbf{z}$) we shift it by a quantity $\epsilon$ as $\textbf{z}_{shift} = \textbf{z} + \epsilon\hat{\textbf{n}}$ along the directions estimated by training SVMs. We use $\textbf{z}_{shift}$ to generate images for different values of $\epsilon$ and find the generated images to reflect the change in the physical parameter encoded by $\hat{\textbf{n}}$ (see Figure~\ref{fig5}). However, physical properties are often entangled, and change in one can cause change in another. As can be observed from Figure~\ref{fig5}, positive shifts along $\hat{\textbf{n}}_{TUF}$ causes an increase in $R$ (new high gradient polarity inversion region) in generated images. This effect can be partially mitigated through conditional manipulation \citep{shen2019}. 
We construct a new direction subtracting the projection along $\hat{\textbf{n}}^{R}$ given by: $\hat{\textbf{n}}^{new} = \frac{\hat{\textbf{n}}^{TUF} - (\hat{\textbf{n}}^{TUF}.\hat{\textbf{n}}^{R})\hat{\textbf{n}}^{R}}{|\hat{\textbf{n}}^{TUF} - (\hat{\textbf{n}}^{TUF}.\hat{\textbf{n}}^R)\hat{\textbf{n}}^{R}|}$
The second row of Figure~\ref{fig5} depicts the change in TUF minimizing the change in $R$. 

For more than one conditional boundary we use orthogonal decomposition of primal direction $x$ onto subspace $W$ constructed by the column space of the matrix $A$, which consists of conditional boundary vectors as columns. The vector perpendicular to the subspace $W$ takes the generic form: $x-A(A^{T}A)^{-1}A^{T}x$.  The third and fourth rows of Figure~\ref{fig5} depict the polarity flip and the change in $R$ with the change in PSEP. The fifth row of Figure~\ref{fig5} shows the effect of conditionally manipulating the change in $PSEP$ over polarity flip and $R$. It shows no polarity flip as $PSEP$ increases and a minimized change in $R$.

\begin{figure}[t!]
\centering
\hspace{-0.03\textwidth}
\includegraphics[width=1.03\linewidth]{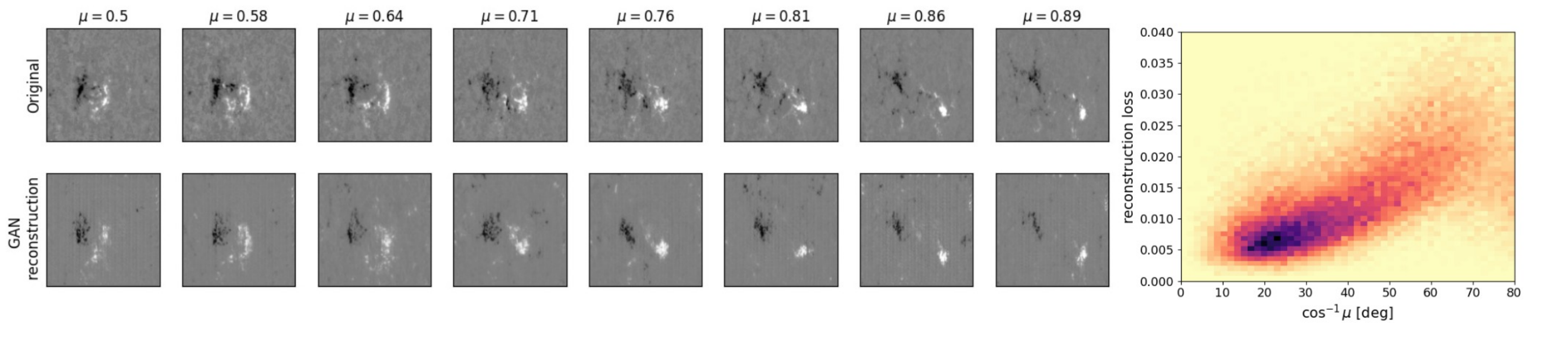}
\caption{Result of GAN approximation (bottom row) of real images (top row) as a function of AR center heliocentric angle ($\cos^{-1}\mu$). The 2D histogram on the right panel shows that the reconstruction loss (i.e. the mean sqaured error between real and GAN approximated images) increase when the AR moves towards the solar limb.}
\label{fig6}
\end{figure}

\textbf{GAN approximation of Real AR image and the effect of AR location.}To estimate the effect of location-dependent noise on GAN outcomes, we approximate the real images with GAN-generated images by tuning the latent vector to minimize the mean-squared error. Mathematically,  it can be described as $\hat{z} = \arg\min_z||Real-G(z)||_2^2$. Plotting $||Real-G(\hat{z})||_2^2$ as a function of the Real AR center heliocentric angle ($\cos^{-1}\mu$), we find that the reconstruction error increases as the AR moves towards the limb (Figure~\ref{fig6}). This increase in reconstruction loss with heliocentric angle happens due to the inability of the GAN to reconstruct the location-dependent noise that is attributed to the dominance of the noisy transverse field at high heliocentric angles. This phenomenon is clearly demonstrated through the comparison of the background noise in original images (upper panels of Figure~\ref{fig6} on the left) with that in GAN-reconstructed images (lower panels of Figure~\ref{fig6} on the left). The background of the original image becomes noisier for higher heliocentric angles, but remains clean for the GAN reconstruction across heliocentric angles. This comparison, quantified by the increasing trend in the loss vs heliocentric angle plot (Figure~\ref{fig6} right-most panel), indicates that the AR location does not affect the GAN-generated images, as the near-limb ARs constitute a small fraction ($<25\%$ for $\cos^{-1}\mu>60^{\circ}$) of the training data. 

\begin{figure}[t!]
\hspace{-0.0\textwidth}
\includegraphics[width=1.0\linewidth]{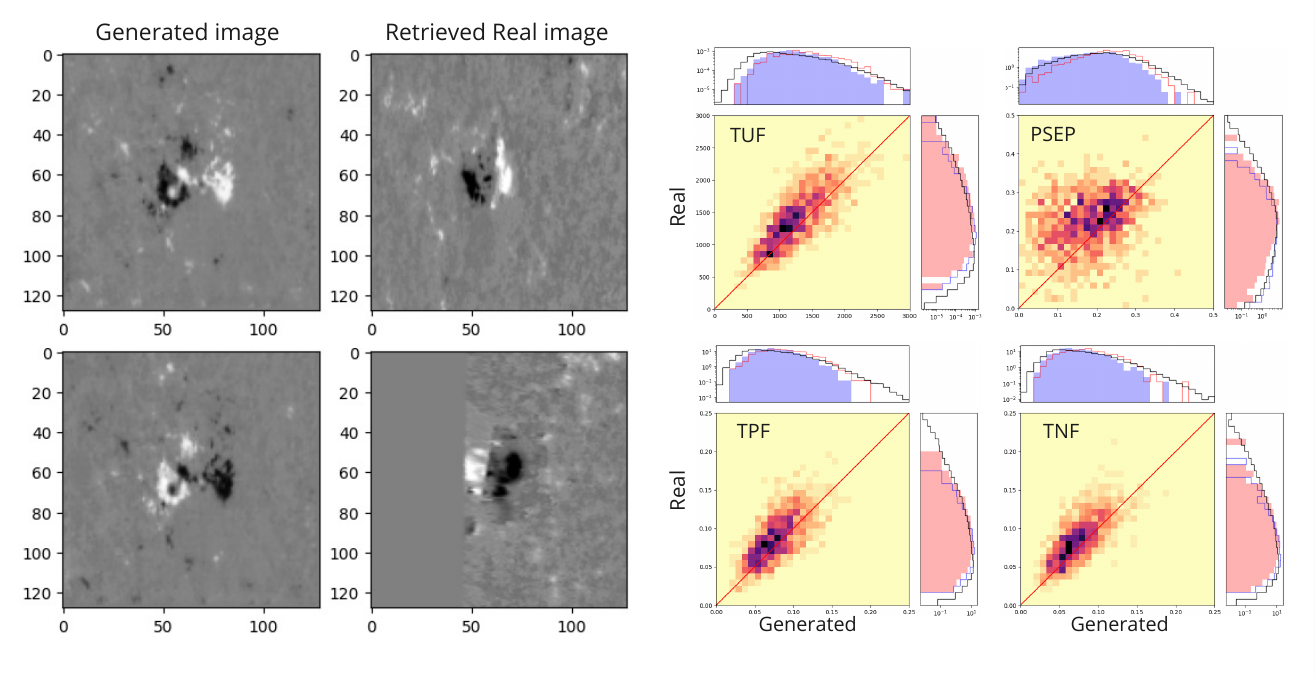}
\caption{\textbf{Retrieval and comparison of physical parameters with real nearest neighbors.} The left-most column shows two generated bipolar active regions with opposite polarity. The 2nd column from the left shows the nearest matching real images to the respective generated queries. The 3rd and 4th columns from the left show the 2D histogram density of the 4 physical parameters (TUF, PSEP, TPF, and TNF) calculated from 1000 randomly generated images and the nearest real neighbors. The marginal histograms of generated and matching real image parameters are plotted in light blue and light pink respectively. The black curves overlaid on the marginals represent the histogram of respective physical parameters evaluated across all real images in the dataset. The marginals for generated images are plotted over real and vice-versa for comparison in respective colors.}
\label{fig7}
\end{figure}

\begin{figure}[t!]
\hspace{-0.0\textwidth}
\includegraphics[width=1.0\linewidth]{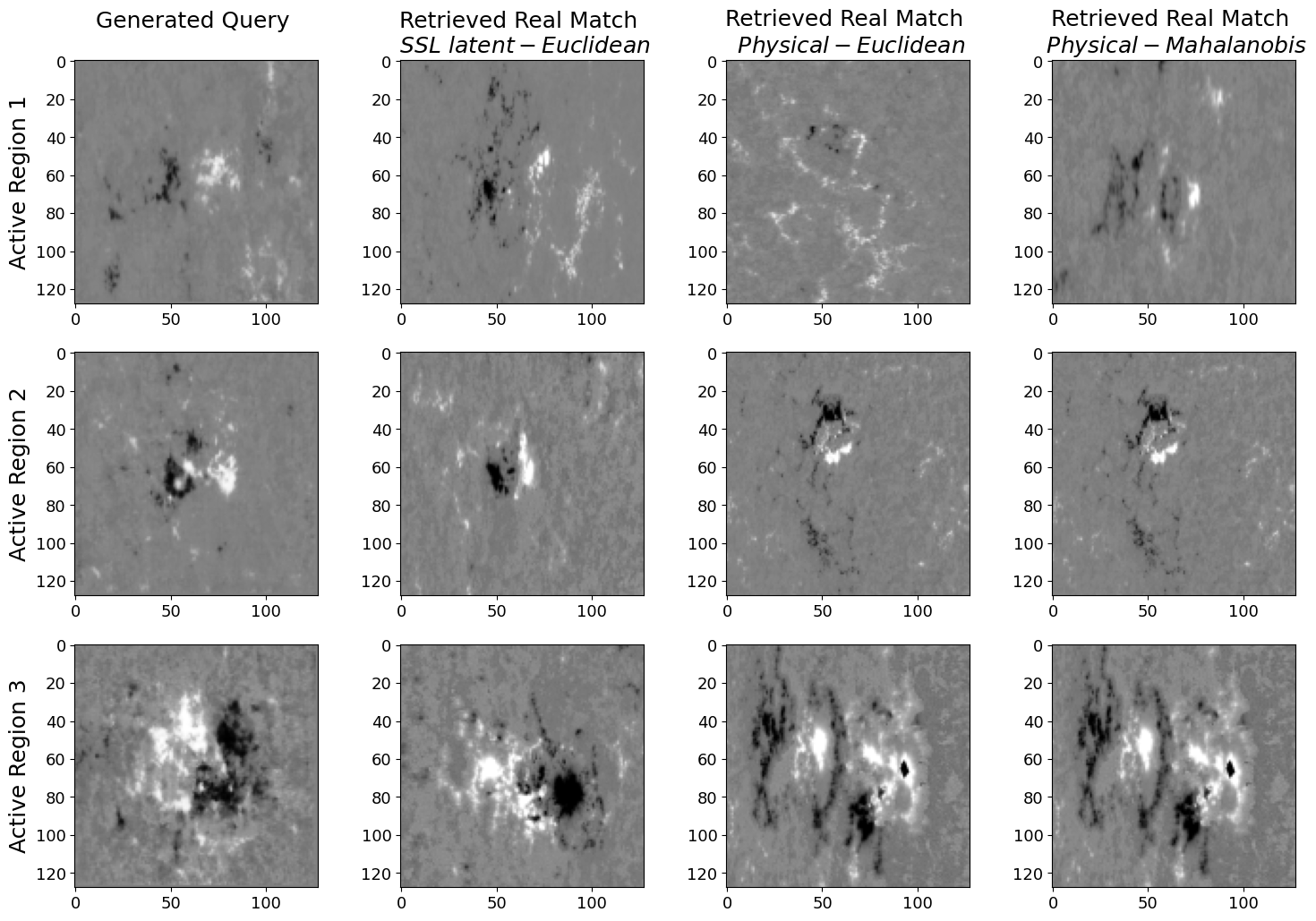}
\caption{\textbf{Comparison of SSL based retrieval of real match with those based on physical parameter space.} Each row shows from left to right, respectively, a synthetic query image, a closest real match based on SSL latent space,  a closest match based on 3-parameter physical space (TUF, PSEP and R) with Euclidean distance, and the one based on the same physical space via Mahalanobis distance.}
\label{fig8}
\end{figure}

\textbf{Querying Generated Image to Find Matches in Real Data.} Although GAN-generated images look realistic, they should possess physical properties comparable with real data. We train a Self-Supervised Learning (SSL) model called SimSiam on the resized SHARPs that learns a 100-dimensional latent space that is invariant across augmentations namely shift (within 10 pixels), zoom (within a factor of 0.8 to 1.2), rotation (within -20$^\circ$ to 20$^\circ$), and vertical flip. The tilt range ensures invariance of latent vectors for an active region appearing at different latitudes and the small zoom range is used to allow for retrieval of slightly different sizes of ARs with similar configuration. It must be noted that those augmentations are optional, and can be removed if a user needs exact match between with query active region in terms of size and tilt. The images and their augmented version are converted to augmentation-invariant latent vectors through two identical CNNs. Once trained, we find the latent representation of the query image and use the Euclidean distance as a metric to find the nearest neighbor in the latent space. The image corresponding to the nearest neighbor in latent space is retrieved as the best match from the set of real images for a query image generated by GAN. We randomly generate 1000 images and retrieve the best-matching real images using the SSL. We compare the distributions of different physical parameters of generated images and real matches. Figure~\ref{fig7} illustrates the retrieval results for a generated query of a bipolar magnetic patch both in its original form and polarity flipped version. The retrieved images seem to conserve the overall polarity arrangement, as well as finer-scale structures. The 2-dimensional histogram of TUF shows a one-to-one correspondence (Pearson correlation $\sim$0.78, Spearman rank correlation$\sim$0.79) between the unsigned field calculated from 1000 generated queries, and corresponding real matches (Figure~\ref{fig7}). The one-dimensional histograms of TUF in Figure~\ref{fig7} depict the match between the generated images, retrieved SHARPs, and the entire space of SHARPs. This validates that our GAN can capture the entire TUF regime of real magnetic patches through its generation and retrieval.  To characterize the correspondence further we calculate the integrated field strength within positive (TPF) and negative (TNF) polarity regions. We observe statistically significant positive Pearson (Spearman) correlations of 0.73 (0.76), 0.28 (0.31), 0.69 (0.7), and 0.72 (0.72)  between generated and closest real images for the parameters $R$, $PSEP$, $TPF$, and $TNF$ respectively. The relatively poor correlation in $PSEP$ comes from the occurrence of the small-scale features and complex multipolar regions that drive the calculation of field-strength-weighted centroids for each polarity. This does not indicate an intrinsic morphological mismatch between the general query and the real match. A better polarity separation measure should cause an improvement in the $PSEP$ correlation. To demonstrate the effectiveness of SSL in image retrieval we compare it with two other much simpler retrieval approaches that use the Euclidean distance and the Mahalanobis distance between physical parameters of synthetic and real data to find matching real ARs. We construct a 3-parameter feature vector ($\textbf{x}$) defined as (TUF, R, PSEP) and calculate the Euclidean distance as $||\textbf{x}_1 - \textbf{x}_2||_2$ and the Mahalanobis distance as $\sqrt{(\textbf{x}_1 - \textbf{x}_2)^T\Sigma^{-1}(\textbf{x}_1 - \textbf{x}_2})$ where $\textbf{x}_1, \textbf{x}_2$ represent feature vector calculated from the synthetic query and real image respectively and $\Sigma$ represents the covariance matrix of real image features. Before calculating the Euclidean distance, we normalize each parameter of the feature vector to avoid unequal contribution; however, that normalization is not necessary for the Mahalanobis distance as it is scale-invariant. Although Mahalanobis distance-based retrieval performs slightly better than the Euclidean one, we find that SSL retrieval ensures a significantly higher visual similarity to the query as compared to other two (Figure~\ref{fig8}). This is likely due to the fact that PSEP, TUF, and R-value do not provide textural properties of images and multiple AR configurations can give rise to same value of those three parameters. Thus in that physical space distance is not strictly indicative of similarity.  But the SSL latent space of dimensionality 100 is more nuanced and meaningful, and is able to address the degeneracy in physical space to image space correspondence. Along with visual match between query and retrieved AR the SSL ensures similarity in terms of physical properties.

\begin{figure}[t!]
\hspace{-0.0\textwidth}
\includegraphics[width=1.0\linewidth]{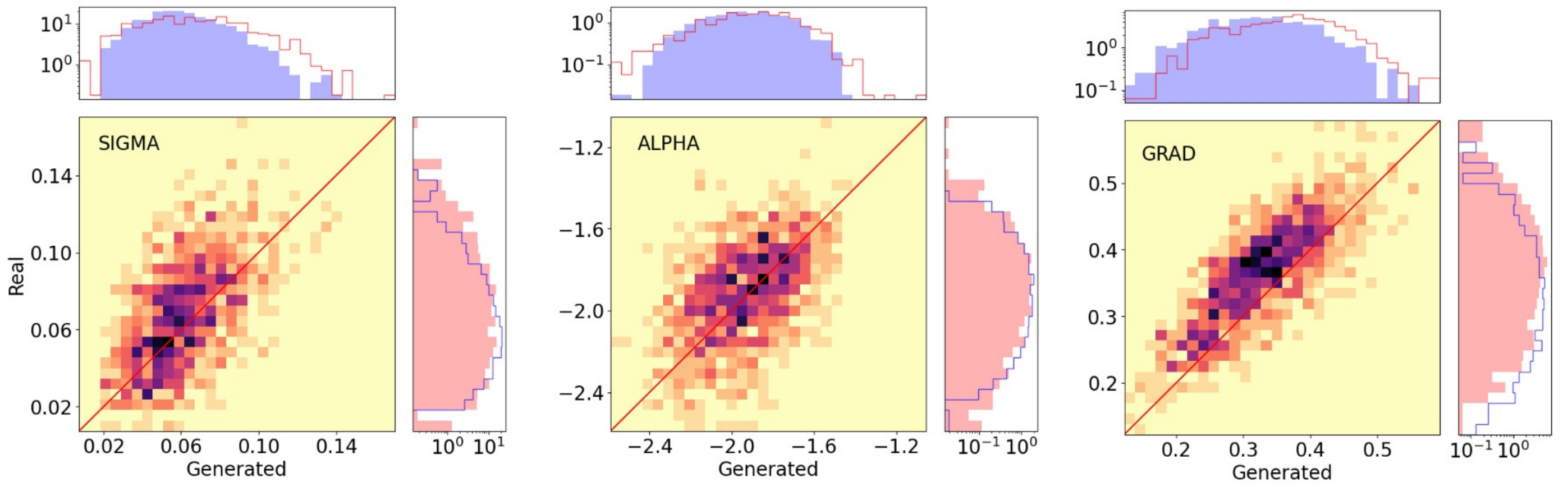}
\caption{\textbf{Comparison of generated query vs real match related to the distributed image features.} The 2D histogram density of the features (SIGMA on the left, ALPHA on the middle, GRAD on the right) calculated from 1000 randomly generated images and the nearest real neighbors. SIGMA represents the standard deviation of a Gaussian probability density function fitted to the histogram density of image intensity. ALPHA represents the slope of the power law fitted to the radially averaged power spectrum of the image for a given spatial frequency range. GRAD represents the average gradient magnitude of an image. The marginal histograms of generated and matching real image parameters are plotted in light blue and light pink, respectively. The marginals for generated images are plotted over real and vice-versa for comparison in respective colors.}
\label{fig9}
\end{figure}

We move beyond the integral image properties discussed previously and calculate properties related to intensity and spatial distribution of image features. We name those properties as SIGMA, ALPHA and GRAD. SIGMA ($\sigma$) is derived from each image by fitting a Gaussian probability density function $\frac{1}{\sigma\sqrt{2\pi}}e^{-x^2/2\sigma^2}$ to the histogram density of image intensity ($x$). ALPHA($\alpha$) is derived by fitting a power-law to the radially averaged image power spectrum $P(f)\propto f^{-\alpha}$ using the first 25 cycles of spatial frequency ($f$). GRAD is defined as the average intensity gradient magnitude ($\sqrt{g_x^2 + g_y^2}$) of an image where $g_x, g_y$ represent $x$ and $y$ gradient at every pixel respectively. We find statistically significant correlation for SIGMA (Pearson$\sim$0.57, Spearman$\sim$0.56), ALPHA (Pearson$\sim$0.47, Spearman$\sim$0.47), and GRAD (Pearson$\sim$0.79, Spearman$\sim$0.78) between generated queries and real matches (Figure~\ref{fig9}).

\newpage
\section{Discussion and Conclusion}\label{sec:discon}
We demonstrated that a Deep Generative model can be harnessed to generate scientifically meaningful queries to find matches in scientific datasets. We achieved this by applying three types of machine learning models --- Generative Adversarial Networks (GANs), Support Vector Machines (SVMs), and a SSL model SimSiam, on the solar magnetic patch dataset, SHARP. Whereas GANs enabled a generation of magnetic active region images from 100-dimensional latent vectors, SVMs helped connect physical parameters (such as total unsigned magnetic field, polarity separation, and total field in the vicinity of polarity inversion lines with high gradients) to the generative latent vectors of GAN. We found that this image generation can be tweaked along directions that reflect changes in the mentioned physical quantities. Depending on the application area, the physical space can be easily expanded under our setup by including other parameters. It will only require the same number of SVM classifiers to be trained for learning latent directions for respective physical properties. It should be noted that an alternative to SVM classifier to connect latent space to physical space would be a multivariate linear regression model that would directly estimate the physical parameters for each point in latent space. But given the higher dimensionality of latent space than the physical space, a given set of physical properties can be satisfied by multiple points in latent space. Those latent space points can give rise to morphologically diverse images with the same physical properties. Thus, an initial latent vector must be modified along the direction normal to the separation boundary to bring the change in a given physical property. The shift along that direction can be calibrated to the value of the physical parameter. As demonstrated in the paper, the directions also help in conditional manipulation, where one requires a change in one physical property without changing another. This would be hard to achieve with a multivariate linear regression setup without considering directions.

We showed that the generated images can also be used as a query to the SSL-derived latent space to retrieve matches from real magnetic regions. These retrieved regions matched the query both visually and quantitatively in terms of integral and spatially resolved image properties. We have demonstrated that a table-search approach using the set of physical parameters as a query cannot ensure visual similarity. The visual similarity pertains to image features that are not described in the physical parameter space. 

There are existing studies that made use of generative models such as conditional GANs (cGAN) to produce synthetic solar data such as time series \citep{Chen_2021} and images \citep{li_2024}. Our work complements those with two key additions-(1) unlike the cGAN approach for conditional generation of synthetic data, our approach doesn’t require retraining the GAN when a condition or physical parameter is added to drive the  image generation. We only need to retrain the SVM classifier with the physical label to learn the corresponding latent direction for image modification. (2) the data generation is not an end goal for our approach. The application of SSL provides a means to connect with the real observations. Our approach, making use of generative and SSL latent space, thus elevates generative models from a means-to-generate-synthetic data to a novel tool for the efficient mining of real scientific data. 

In the domain of Heliophysics, this approach can serve as a generic framework for solving problems such as instrument-to-instrument translation, artifact correction and reconstruction of farside active regions. All those transformations can be learned as directions in latent space along which a real image can be modified to meet desired characteristics related to a different instrument or occurrence of artifacts or physical properties. For space weather forecasting it is important to offset the class imbalance problem by adding synthetic examples to minority classes \citep{Chen_2021}. Our approach would help create ARs that lead to flares, Coronal Mass Ejections, or Solar Energetic Particle events by learning latent space directions. Modifying ARs along those directions would increase the likelihood of them causing space weather events. The SSL module will additionally fetch real ARs with matching morphology and physical characteristics, serving as a validation framework.
In addition to the domain of heliophysics this approach can be easily adapted to any other field of astronomy dealing with big datasets of any modality and complexity. In the future, we would like to improve our software to use numerical values of physical parameters (e.g., calibrate the change in magnetic-flux-influencing parameter to the value in actual physical units in Mx). This will enable more precise control of the generation of queries. 
\vspace{0.1\textwidth}
\section{Data and Code Availability} \label{sec:cite}
The SHARPs, used to train the generative model for this study, are available from the Joint Science Operations Center
(\url{http://jsoc.stanford.edu}).

The codebase we developed for this work can be accessed from \url{https://doi.org/10.5281/zenodo.17613896}\citep{chatterjee_2025_genai}.

\section*{acknowledgments}
This research was funded by NASA HGIO grant 80NSSC23K0416.

\bibliography{references}{}
\bibliographystyle{aasjournal}

\end{document}